\pdfoutput=1
\documentclass[epj]{svjour}

\usepackage[usenames]{color}
\usepackage{chemfig,bm,rotating}
\usepackage{float}
\usepackage{graphicx}
\usepackage{ulem}

\begin{document}

\title{Multiplasmon excitations in electron spectra of small systems 
irradiated by swift charged projectiles}
\titlerunning{Multiplasmon excitations in spectra of systems 
irradiated by swift charged projectiles}
\author{P. M. Dinh\inst{1,2}\fnmsep\thanks{\email{dinh@irsamc.ups-tlse.fr}}
\and
P.-G. Reinhard\inst{3}
\and 
E. Suraud\inst{1,2}
\and
P.~Wopperer\inst{1,2}
} 

\institute{
Universit\'e de Toulouse; UPS; Laboratoire de Physique Th\'eorique
  (IRSAMC); F-31062 Toulouse, France
\and
CNRS; LPT (IRSAMC); F-31062 Toulouse, France
\and
Institut f\"ur Theoretische Physik, Universit\"at
Erlangen, D-91058 Erlangen, Germany
}

\date{\today}

\abstract{
We investigate the kinetic-energy spectrum of electrons emitted
  from an excited many-electron system, often called photo-electron
  spectrum (PES).  We are particularly interested on the impact of
  resonant modes of the system on PES. To this end, we consider three
  systems with strong resonances, a Mg atom, the small alkaline cluster ${{\rm K}_9}^+$,
  and the small carbon chain C$_3$.  To avoid dominant frequencies in
  the excitation process, we consider a collision with a fast ion
  which is realized by an instantaneous boost of the valence
  electrons, a process which excites all frequencies with equal
  weight. The electron dynamics is investigated from a theoretical
  perspective using time-dependent density-functional theory augmented
  by an average-density self-interaction correction. We observe
patterns which are similar to PES usually obtained after irradiation by
a laser pulse, in particular the appearance of clear peaks. We show
that these patterns are driven by strong resonance modes of
the system. Resonances are thus found to be another source of 
peaks in the PES, besides photons (when present) with definite frequencies.
}

\maketitle

\section{Introduction}
\label{sec:intro}

Photo-Electron Spectra (PES) are a widely used and powerful
observable to analyze electronic systems \cite{Ber79,Gho83}. The
conceptually simplest setup is to use a photon pulse with a well defined
frequency which is sufficiently high in energy to ionize a great amount of
single electron states by a one-photon excitation. This provides a
direct map of the single-particle (s.p.) energies, for examples from
cluster physics, see \cite{Leo87,Kos05}.  More involved is the
interpretation in the regime of Multi-Photon Ionization (MPI) where
PES can contain multiple copies of the s.p. spectra according to
different photon numbers, often overlaid by thermal
emission \cite{Cam00}. As a further contributor to PES, one often
finds traces of the pronounced surface plasmon modes. Plasmon
satellites are observed experimentally in X-ray PES of bulk
metals~\cite{Gro06}, metal clusters~\cite{And11,And12}, nested
fullerenes~\cite{McC11}, and C$_{60}$~\cite{Mau09}. A competition
between photon and plasmon peaks in metal clusters was also worked out
theoretically for MPI with typical laser frequencies in \cite{Poh01}. It
is, however, not always easy in the PES to disentangle peaks due
  to plasmon modes, or other strong resonances, from photon
  signals. In order to distill the effect of resonant modes, we
    choose an excitation which in itself has no frequency bias.  This
  is achieved by very fast electro-magnetic pulses as they are
  delivered by Coulomb collisions with fast ions. We model them
    in practice by an instantaneous boost of all valence electrons
    which excites all frequencies with equal weight, such that
  pronounced frequencies in the PES can only stem from the
  target molecule. This allows one to see the impact of strong modes
  on the PES more unambiguously than in the case of a laser
  excitation.

Collision experiments are still comparatively rare. None\-the\-less, there
exist already a few promising measurements which resolve properties of
the emitted electrons, e.g. for collision of protons on uracil
delivering total cross-sections and PAD~\cite{Agn12,Tri12} as well as
PES~\cite{LeP08}.  These experiments are still lacking proper
theoretical analysis, but they will certainly bring invaluable
information on dynamical mechanisms by such a detailed analysis of the
properties of emitted electrons. At present, even a theoretical
proof of principle would bring a helpful piece of information. Thus it
is the aim of this paper to investigate the impact of resonant
  modes on PES from a theoretical side. The fast collisional process
  is modeled in a simple manner by an instantaneous dipole boost of the
  electron cloud.  The possible availability of more collision
experiments is a strong motivation for the present paper.  As test
cases, we consider three rather different systems, that is, 
the Mg atom, the
metal cluster ${{\rm K}_9}^+$, and the C$_3$ chain.
They however possess the common feature of being sufficiently
metallic in the sense that their optical spectra are dominated by collective modes, 
as will be discussed at length for each case.

The paper is outlined as follows: The theoretical and numerical
background is sketched in section \ref{sec:theory}. Results for the Mg
atom are discussed in section \ref{sec:mg}, for the ${{\rm K}_9}^+$
cluster in section \ref{sec:alkaline}, and for the C$_3$ chain in
section \ref{sec:Cchains}. We finally draw some conclusion in the last
section.

\section{Theory and numerical scheme}
\label{sec:theory}

\subsection{Dynamical simulation}

We describe the electronic dynamics by means of real-time Time-Dependent Density
Functional Theory (TDDFT) in a standard manner~\cite{Cal00,Rei03a}.  We
solve the (time-dependent) Kohn-Sham equations for the cluster
electrons on a grid in coordinate space, using time-splitting for time
propagation~\cite{Fei82} and accelerated gradient iterations for the
stationary solution \cite{Blu92}. The Poisson equation is solved by a
fast Fourier technique combined with separate treatment of the
long-range terms~\cite{Lau94}.  We use the exchange-correlation
energy functional from Perdew and Wang~\cite{Per92}.  A
Self-Interaction Correction (SIC) has to be applied to obtain
correct s.p. energies which are crucial for an
appropriate dynamical description of electron emission. We
include it by the technically inexpensive Average-Density SIC
(ADSIC)~\cite{Leg02}. The coupling to the ions is mediated by soft
local pseudopotentials in the case of Mg and K~\cite{Kue99}, and
Goedecker-like ones in the case of C~\cite{Goe96}.  We use absorbing
boundary conditions~\cite{Cal00,Rei03a,Rei06c} which gently absorb all
outgoing electron flow reaching the bounds of the grid and thus
prevent artifacts from reflection back into the reaction zone.

Calculations for the Mg atom were done on a cylindrical
grid extending 110 a$_0$ along the $z$ axis and 55 a$_0$ in
$r$ direction, with a mesh size of 0.5~a$_0$. The ${{\rm K}_9}^+$ cluster is also treated in cylindrical
symmetry in a box of mesh size of 0.9~a$_0$, and of length 200 a$_0$ in $z$ and 100 a$_0$ in $r$.
In this case, the cylindrical symmetry for the Kohn-Sham potential is an
approximation, the cylindrically averaged pseudopotential scheme
\cite{Mon94a,Mon95a}, which has proven to be an efficient and reliable
approximation for metal clusters close to axial symmetry.
The C$_3$ chain is computed on a three-dimensional grid with box
lengths of 70 a$_0$ in each spatial direction, and a mesh size of 0.7~a$_0$.

\subsection{Ionic collisions and instantaneous boost}

The excitation mechanism used throughout this paper is an
instantaneous dipole boost which is applied to all occupied single
electron states $j$. For example, a boost by $p_z$ in $z$-direction
reads
$\varphi_j(\mathbf{r},t\!=\!0)=\exp(\mathrm{i}p_zz)\varphi_{j,\mathrm{gs}}(\mathbf{r})$
where $\varphi_{j,\mathrm{gs}}$ are the ground-state wave
functions. This approximately models the effect of the Coulomb field
of a fast charged projectile passing by.  A sufficiently fast ion
  moves near the molecule practically on a straight line with constant
  velocity $v$. Let us assume that the ion moves orthogonal to the $z$
  axis in a frame whose origin lies at the molecule's center. The point
  of closest approach then lies on the $z$ axis at a distance $b$,
  identical with the impact parameter. The Coulomb field of the ion
  exerts a force on the molecule electrons. Integrating the force
  over the collision process, we find that the net force has only a
  component in $z$ direction \cite{Bae06a}. For a sufficiently remote
 impact parameter $b$, the force field at the molecule site is practically homogeneous
  and boosts the electron momentum by
\begin{equation}
  p_z
  =
  4Ze^2/(bv)
\label{eq:pboost}
\end{equation}
where $Z$ the charge of the projectile and $e$ the electron charge.
Let us consider typical orders of magnitude for the maximally amenable
boost momentum $p_z$. One condition is that the ionic velocity has to
be large such that the passage is much shorter than the electronic
reaction time. The time for the passage is of order $\delta
t=b/v$. This has to be shorter than the relevant reaction times
$\omega_\mathrm{el}^{-1}$.  Thus we require
$v/b\gg\omega_\mathrm{el}$. For a rough estimate, we assume
$\omega_\mathrm{el}=1$ Ry and an impact parameter $b=10$ a$_0$ which
should stay outside the cluster's electron cloud. The limiting value
for $v$ is then
$v=b\omega_\mathrm{el}=10\,\mathrm{Ry}\,\mathrm{a}_0=200\,\mathrm{a}_0/\mathrm{fs}$.
For a colliding proton, this corresponds to $E_\mathrm{kin,p}=0.7$
MeV.  More interesting is the corresponding boost strength.  Inserting
the $b=10$ a$_0$ and $v=10\,\mathrm{Ry}\,\mathrm{a}_0$ 
into the above formula for $p_z$ yields $\delta
p_z=0.08Z/$a$_0$. With $Z=1$ (proton), this is in the range of boosts used later.  The
estimate is at the optimistic side. One will probably need somewhat
larger $b$ and $v$. This reduces the achievable $p_z$.  However,
  there is still the option to consider ions with higher charge states
  $Z$. This allows one to tune the boost strength in a wider range.  In the
  following, we will work out a range of boost strengths where
  resonant effects on PES are visible.

\subsection{Extracting observables}

Optical response, that is, the photo-absorption strength distribution,
constitutes the basic information on the collective modes and the
particle-hole excitations of a system.  We will compute the
  dipole strength distribution tracking the dipole signal following a
faint instantaneous boost and subsequent spectral analysis,
i.e. Fourier transforming the time-evolution of the dipole moment into
the frequency domain~\cite{Cal97b}.

The observable in the focus of this paper is the kinetic-energy
spectrum of electrons emitted after the boost. In analogy to laser
experiments, we call that a photo-electron spectrum (PES).  It is
computed in the same simple fashion as done in the first studies of
laser excitations~\cite{Poh00}. Note that strong and/or long laser
pulses require more elaborate techniques~\cite{Din13} which are
fortunately ignorable for the instantaneous excitations considered
here. We choose a couple of ``measuring points''
$\mathbf{r}_\mathcal{M}$ far away from the emitting system and just
before the absorbing boundaries.  Out there, the Kohn-Sham field can
be considered negligible and a free-particle dynamics can be assumed.
Moreover, the closeness of the absorbing bounds allows one to assume
that only outgoing waves with momentum
$\mathbf{k}=k\mathbf{r}_\mathcal{M}/r_\mathcal{M}$ and $k>0$ pass the
point $\mathbf{r}_\mathcal{M}$.  We now record a protocol of each
s.p. wave functions $\varphi_j(\mathbf{r}_\mathcal{M},t)$ during
simulation.  As we encounter practically outgoing free waves at
the $\mathbf{r}_\mathcal{M}$, the kinetic-energy content of the wave
function is equivalently contained in the frequency spectrum of the
wave functions.  The PES yield
$\mathcal{Y}_{\Omega_{\mathbf{r}_\mathcal{M}}}(E_{\rm kin})$ can thus
be obtained from Fourier transformation from time to frequency
$\omega$ of $\varphi_j$ as
\begin{equation}
  \mathcal{Y}_{\Omega_{\mathbf{r}_\mathcal{M}}}(E_{\rm kin}) 
  \, \propto \, 
  \sum_{j=1}^N
  \left| \widetilde{\varphi}_j(\mathbf{r}_{\mathcal{M}},E_{\rm kin})\right|^2   
  \,,
\label{eq:raw3D}
\end{equation}
where $\Omega_{\mathbf{r}_\mathcal{M}}$ is the solid angle related to
the direction of $\mathbf{r}_\mathcal{M}$, and
$\widetilde\varphi_j(\mathbf{r}_\mathcal{M},E_\mathrm{kin})$ is the
time-frequency Fourier transform of the wave function of state $j$,
$\varphi_j(\mathbf{r}_\mathcal{M},t)$, with $E_{\rm kin}=k^2/2m_e=
=\omega$~\cite{Poh00,Din13}.  Each term in the sum of Eq.~(\ref{eq:raw3D})
represents the PES for emission from a single state $j$. Such
a state-specific PES can be used for analyzing the results. Moreover, taking a
sufficiently fine mesh of measuring points $\mathbf{r}_\mathcal{M}$
allows one to evaluate the angular-resolved PES. 
In the present paper, we will mostly consider angular-averaged PES, i.e., the
PES according to Eq.~(\ref{eq:raw3D}) averaged over all
$\mathbf{r}_\mathcal{M}$ with appropriate angular weights, and discuss
an energy- and angle-resolved PES only in the case of C$_3$.
Although the boost is a fast excitation, electron emission and
accumulation of PES takes a while because slow electrons reach the
measuring points late. Computations were carried forth
long enough to cover a large fraction of emitted electrons, from 64 fs
for the C chain up to 300 fs for the K cluster.

\section{A simple example~: the Mg atom}
\label{sec:mg}

\begin{figure}[htbp]
\begin{center}
\includegraphics[width=\linewidth]{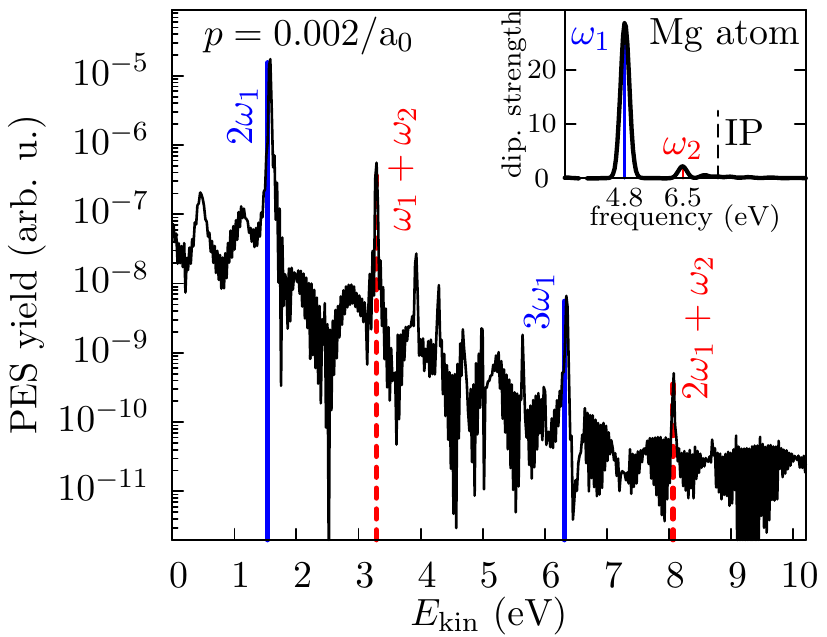}
 \caption{\label{fig:mg_atom}
Calculated photoelectron spectrum of the Mg atom after an excitation by
an instantaneous boost of 0.002/a$_0$.  The vertical lines emphasize the 
kinetic energies as given by Eq.~(\ref{eq:multiplas}).
Inset~: dipole strength (in vertical linear scale and
arbitrary units) with two dominant peaks located at $\omega_1=4.8$~eV and 
$\omega_2=6.5$~eV. The dashed line indicates the value of the ionization potential
at 8~eV.
}
\end{center}
\end{figure}

As a starter, we choose the simple and oversee-able example of a
single Mg atom to investigate the effect of dominant excitation modes
on the PES after instantaneous excitation. The Mg atom is ideal for our
purposes because it has a single valence state ($E_{\rm IP}=7.6$~eV
experimentally~\cite{Lid08}, 8 eV in our calculations) and it has a strong
$2s\longrightarrow 2p$ dipole transition which dominates the spectrum
by far. This can be seen from the photoabsorption spectrum in the
inset of Figure \ref{fig:mg_atom}, with the peak labeled
$\omega_1$.
One can, however, notice further smaller excitation states following
at higher energies. Among them, the second transition, marked by $\omega_2$, has
still a sizable strength, and we will see in the following that it
does play a role in the PES. Note also that most of the optical response
of the Mg atom lies below the ionization potential (IP).

In spite of the fact that the instantaneous boost excitation covers a
continuous spectrum of frequencies, the PES clearly shows distinctive
peaks above background, as cane seen from Figure \ref{fig:mg_atom}.
The kinetic energies
$E_\mathrm{kin}$ at which the most prominent peaks of the
PES occur can be associated with the strongly excited states according
to the rule~:
\begin{equation}
  E_\mathrm{kin}
  =
  \varepsilon_j+ \sum_{k=1}^{M} \nu_k\omega_k \quad,
\label{eq:multiplas}
\end{equation}
where $\varepsilon_j$ denotes the s.p. energy of orbital $j$, the
integer $k$ runs over the $M$ various strong modes inferred from the
optical response, and $\nu_k$ are integers.  This looks 
  similar to the rule $E_{\rm kin}=\varepsilon_j+\nu\omega_{\rm las}$
  which applies if the system is excited by a laser pulse with
  frequency $\omega_{\rm las}$.  The difference is that the photon
  frequency $\omega_{\rm las}$ is imposed from outside the system
  while here in Eq.~(\ref{eq:multiplas}) the dominant frequencies are
  system properties. In the situation displayed in
Figure~\ref{fig:mg_atom}, we have for the Mg atom $M=2$ because there 
are 2 dominant peaks in the optical response (see inset), and the
coefficients $\nu_1$ and $\nu_2$ can take values from 0 to 3 (for instance,
the last peak indicated by a vertical line in the PES corresponds to $\nu_1=2$ and
$\nu_2=1$). As
expected, the peak height in the PES is related to the strength of
each mode~: the highest peaks are associated to a double and a triple
plasmon excitation of $\omega_1$, while the peaks raising from
linear combinations of $\omega_1$ and $\omega_2$ show smaller relative
heights, either in the case of a double plasmon excitation or a triple
one.

One can even further notice a couple of more peaks of smaller
strength. They may be associated with the smaller peaks in the
photo-absorption spectrum. But a detailed assignment according to
Eq.~(\ref{eq:multiplas}) becomes soon untraceable because there are
too many possible combinations of the various minor peaks. This
example thus nicely demonstrates that strong excitations can be seen
as peaks in the PES and that one needs rather dominant and distinct
excitation modes for an unambiguous analysis.

\section{The case of an alkaline cluster}
\label{sec:alkaline}

\begin{figure*}[htbp]
\begin{center}
 \includegraphics[width=0.8\linewidth]{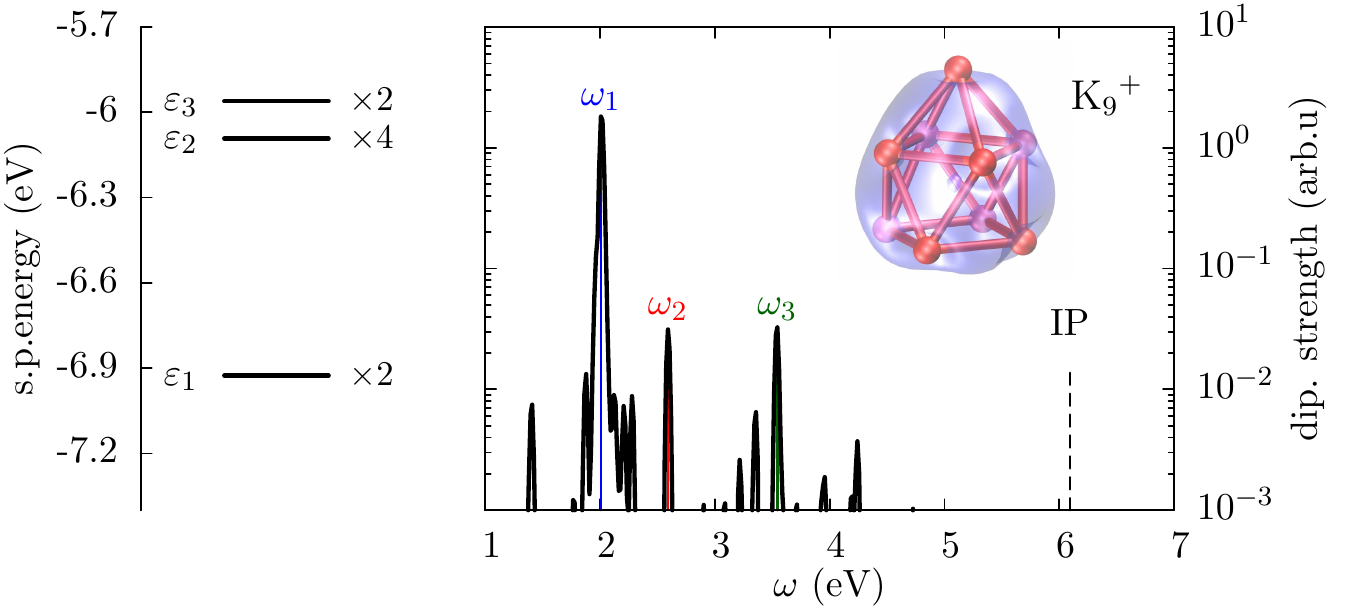}
 \caption{\label{fig:K9p-optresp} 
Left~: single particle (s.p.) energies of ${{\rm K}_9}^+$ with degeneracies indicated at the right
side of each level. Right~: calculated optical response of ${{\rm K}_9}^+$, with the three dominant
frequencies labeled as $\omega_{1,2,3}$ and the ionization potential indicated by 
the vertical dashed line. The electronic density and the ionic structure of ${{\rm K}_9}^+$ appear
as an inset.
}
\end{center}
\end{figure*}

Metal clusters are known for their prominent Mie surface plasmon
\cite{Kre93}. However, this plasmon can be strongly fragmented due to
  interference with s.p. excitations in the case of heavier clusters
\cite{Rei96b}. Small clusters offer the best chances to encounter a
clean excitation. Figure \ref{fig:K9p-optresp} shows the dipole
excitation spectrum (in logarithmic scale) of the ${{\rm K}_9}^+$
cluster.
The excitation spectrum is indeed dominated by the Mie surface plasmon
associated here with the frequency denoted $\omega_1$. But there is a
couple of other dipole modes. The next two important ones are labeled
by frequencies $\omega_2$ and $\omega_3$. We finally note that, as in
the case of the Mg atom, the IP of ${{\rm K}_9}^+$, at 6.1~eV, 
lies much above its optical response.

\begin{figure}[htbp]
\begin{center}
 \includegraphics[width=\linewidth]{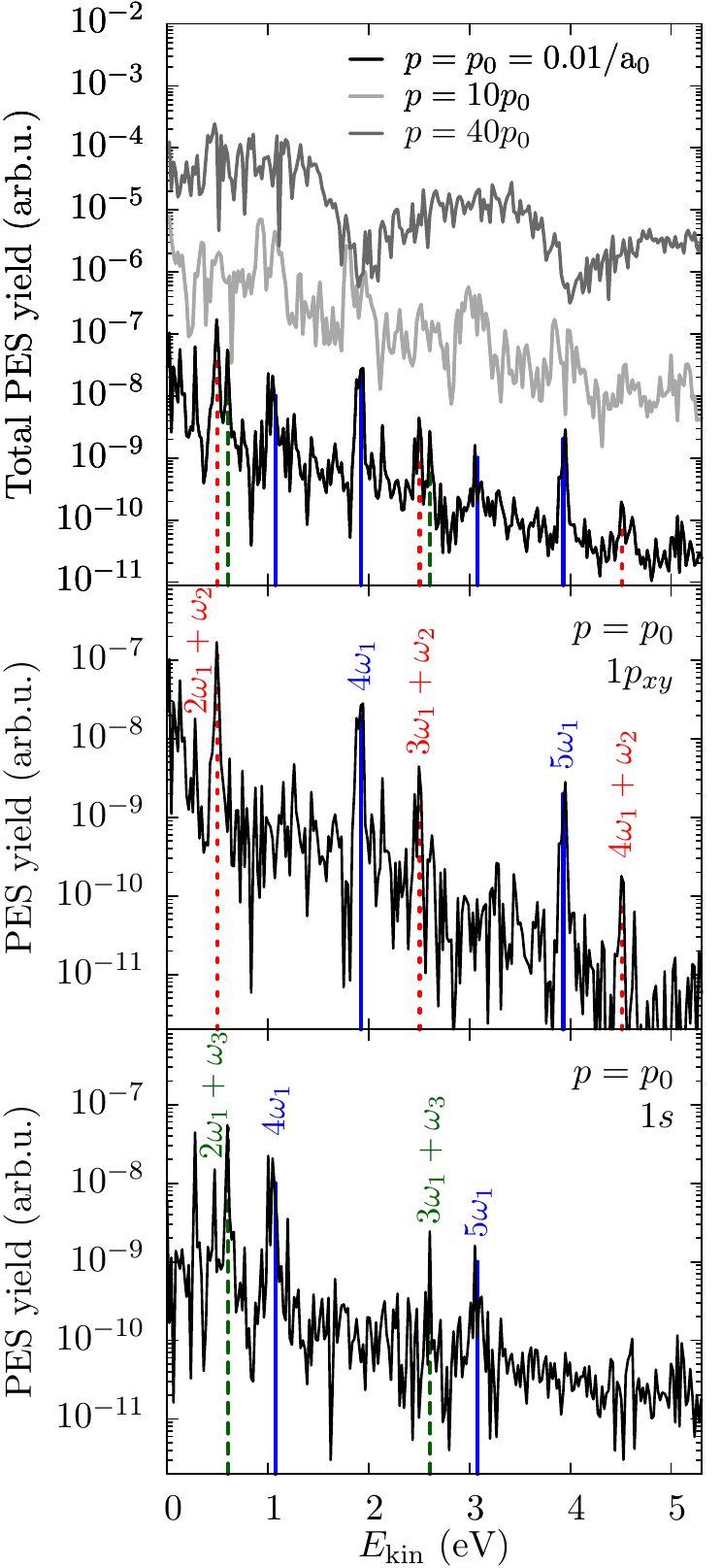}
 \caption{\label{fig:K9p-plasmon}
   Photoelectron spectra (PES) from instantaneous boost  for a ${{\rm K}_9}^+$ cluster.
   Upper: Total PES for three
   boost strengths as indicated. 
   The PES are rescaled to achieve a better visibility.
   Middle: 
   PES from the  $1p_{xy}$ state for boost momentum $p=p_0=0.01/$a$_0$.
   Lower:
   PES from the  $1s$ state for boost momentum $p=p_0=0.01/$a$_0$.
}
\end{center}
\end{figure}

We now turn to the PES obtained after an instantaneous boost, see
Figure~\ref{fig:K9p-plasmon}.
The upper panel shows the PES obtained with three different
instantaneous boost momenta $p$ as indicated.  The total number of
emitted electrons increases from 0.007 for the lowest boost $p=p_0=0.01$/a$_0$,
up to 0.54 for $p=40p_0$. Correspondingly, the peak structure gets
more and more blurred with a steady downshift and a broadening of the
peaks. As in previous studies~\cite{Poh01}, a total electronic
emission of $\approx 1$ seems to be the critical ionization where PES
peaks become completely dissolved.  This means that a clean
signal can be obtained only in a very specific range of impact
parameters: The boost should be large enough to raise
the signal above background, but also small enough to avoid washing
out the PES.

We thus discuss in more detail the smallest of the three boost
excitations, since it delivers the cleanest signal~: It yields a very small ionization which leaves the
s.p. energies practically unchanged (the problem of the
Coulomb shift from ionization and its impact on PES will be discussed in the next
section). The two lower panels of Figure~\ref{fig:K9p-plasmon} show
state-specific PES. The system has three groups of (degenerate)
states, a deep lying twofold $1s$ state at
$\varepsilon_{1s}=-6.9$~eV, and a slightly splitted group of $1p$ states
with fourfold $\varepsilon_{1p_{xy}}=-6.1$~eV and twofold
$\varepsilon_{1p_z}=-6.0$~eV.  Emission from the $1p_z$ state is
suppressed as compared to the two other states due to unfavorable
coupling matrix elements. The figure thus shows only the two relevant
contributions to the PES, the middle panel from $1p_{xy}$ and the
lower panel from $1s$. The total PES is then the superposition of
the two separate spectra.

Both PES show a variety of peaks which look at first glance rather
puzzling. Of course, a cluster is more complex than a simple atom.
Consequently, the PES is more involved.  The various vertical 
lines in the plot indicate frequencies according to the rule
Eq. (\ref{eq:multiplas}) with $\varepsilon_j=\varepsilon_{1p_{xy}}$ in the
middle panel and $\varepsilon_j=\varepsilon_{1s}$ in the lower one.
We see that all larger peaks can uniquely be associated to a proper
mix of excitation frequencies. The simplest way to explain peaks are
those which contain just a multiple of the plasmon frequency
$\omega_1$ (see blue solid lines); they are indeed found to be prominent peaks in the total
PES in the top panel of the figure, for a given ensemble of multi-mode excitation 
(4-plasmon excitation, 5-plasmon one, etc.).  Apparently, the energy given by the boost to the cluster is
first mainly stored in the collective plasmon and then, from time to
time, transferred to single electrons. This is revealed, e.g. in the 4-plasmon excitation, by a peak
at $\varepsilon_{1p_{xy}}+3\omega_1+\omega_2$ in the middle panel, and a 
peak at $\varepsilon_{1s}+3\omega_1+\omega_3$ in the bottom panel. Note also
that for the 3-plasmon excitation, $3\omega_1$ is not sufficient to ionize the $1p_{xy}$ state.
This is why we rather assign the peaks at low $E_{\rm kin}$ to excitations from $2\omega_1+\omega_2$
or $2\omega_1+\omega_3$.

This example on ${{\rm K}_9}^+$ thus nicely demonstrates that the
actual excitation spectrum has a direct impact on the PES.  However,
it also indicates that a metal cluster already produces a complex
picture and it may not be straightforward to recover clear plasmon
signals in collision experiments where we usually do not have
the possibility to disentangle contributions from the separate states.

\section{The C$_3$ chain}
\label{sec:Cchains}

We now turn to the C$_3$ chain as a more complex system which is
initially built on covalent binding, but which also exhibits
partially metallic behaviour. Small
carbon molecules can be produced in laboratory by thermal/laser
vaporization of graphite or by electron impact induced fragmentation
of hydrocarbons \cite{Roh84,Lif00}.  Among all physical properties,
their structure has been under debate since long.
Electron affinities measured by photoelectron spectroscopy as well as
abundances show an even-odd alternation for size below 20
\cite{Wel89,Arn91,Yan88}.  This supports the assumption that linear
chains are the predominant structure at these sizes since chains with
odd numbers of atoms have closed-shell ground states, while chains with
an even number of atoms have
open-shell ones.  For the purpose of this study, we focus on the
smallest carbon chain, that is C$_3$.  In the following, we first
discuss the static properties of C$_3$ (energy spectrum and optical
response) and then move to its PES produced by an instantaneous boost.

\subsection{Static properties and optical response}
\label{sec:c3-stat}

\begin{figure*}[htbp]
 \begin{center}
\includegraphics[width=0.8\linewidth]{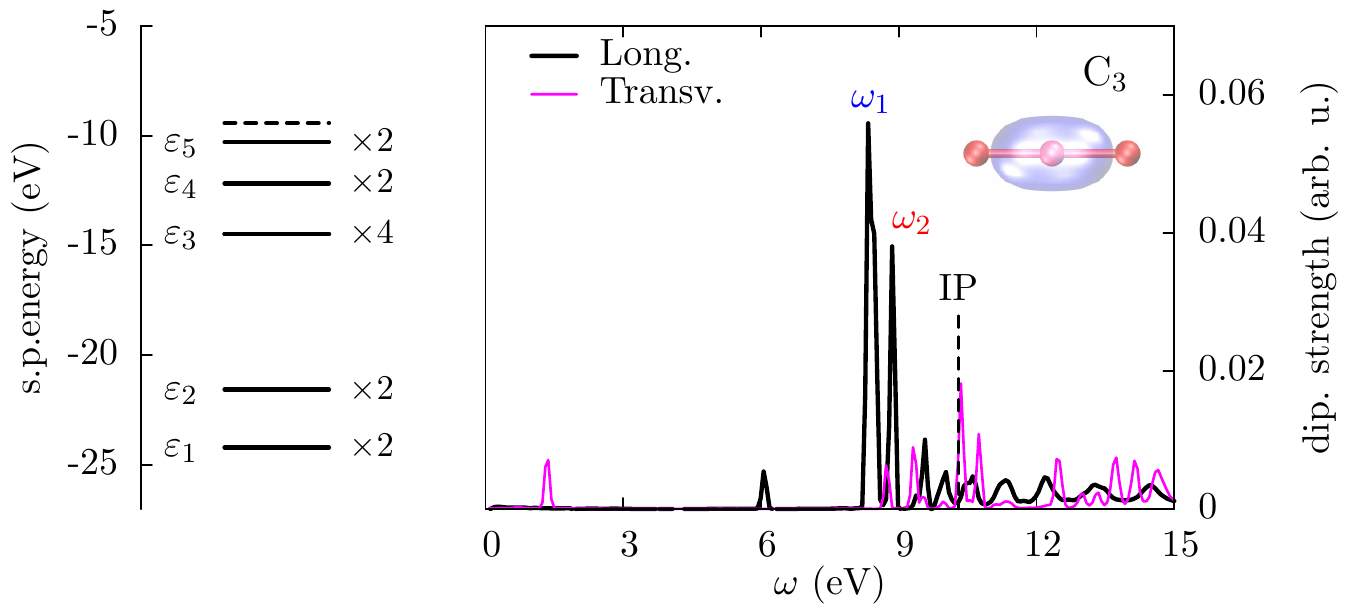}
 \caption{Basic properties of the C$_3$ chain. Left~: calculated single
   particle spectrum with degeneracies indicated at the right side of
   the levels, the numbering of states at the left. The dashed line represents the LUMO. 
Right~: calculated
   optical response with the two strongest longitudinal modes associated with frequency labels
   $\omega_{1,2}$ and the position of the ionization potential (IP)
   indicated by a vertical dashed line. The ionic structure of C$_3$ with the electronic density
appears as an inset.}
 \label{fig:c3_stat-opt}
\end{center}
\end{figure*}

The C$_3$ chain is a simple linear molecule.  The C-C bond length of
2.414~a$_0$ used in our calculations agrees with previous theoretical
results (between 2.415 and
2.485~a$_0$)~\cite{Kos08,Zha02,Men93,Xu92,Rag87}.  The s.p.
spectrum of C$_3$ as well as the degeneracy and labeling of each state
is displayed in the left panel of Figure~\ref{fig:c3_stat-opt}.  The
C$_3$ has 12 valence electrons which group into 5 (degenerated) states
with a large gap of about 7~eV between the two deepest levels and the
three higher ones.
Our calculated value of the IP for C$_3$,
$E_\mathrm{IP}=-\varepsilon_5=10.8$~eV, is in fair agreement with
other calculations (12.0~eV~\cite{Dia02,Lia08}) and experimental
measurements (11.6~eV~\cite{Bel07}, 12.1~eV~\cite{Ben05}, and
13.0~eV~\cite{Ram93}).  
Our calculated IP is a bit too low. This is because we refitted the parameters of the
Goedecker-like pseudopotential to allow us to use a mesh size of 0.7~a$_0$.
If we use the original value of the Goedecker pseudopotential~\cite{Goe96}, we obtain an
IP of 11.3~eV~\cite{Klu13}, much closer to the experimental value. However, that would mean a mesh size
twice smaller (0.36~a$_0$), that is huge numerical boxes which represent a great hindrance for the
dynamical calculations we present here. 
So, there is indeed a dependence of the IP on the pseudopotential parameters
that we use. However, too low an IP does not significantly impact the principle effects
investigated here.

We now turn to the photoabsorption strength of C$_3$ shown on the
right part of Figure~\ref{fig:c3_stat-opt}. In contrast to the Mg atom
and the ${{\rm K}_9}^+$ cluster which are more or less spherically
symmetric systems, the carbon chain looks and behaves very different
in the longitudinal direction (along the symmetry axis = $z$ axis) and the
transverse one. The response in the transverse direction is
suppressed with respect to that in the longitudinal direction. This is
not surprising because the restoring forces are stronger for
transverse modes, thus forcing smaller amplitudes.  We also observe a
strong spectral fragmentation in both directions. The
strongest longitudinal mode resides at $\omega_1=8.4$~eV, closely
followed (in position and strength) by another strong longitudinal
mode at $\omega_2=8.8$~eV. These two modes together exhaust the
dominant fraction of the longitudinal dipole strength. We expect that
they should produce visible effects in the PES.  The
  resonance frequency is in very good agreement with previous
theoretical calculations ($\omega_1=8.1$~eV~\cite{Yab97,Kol95}).
However, all theoretical results lie much higher than the measured
plasmons~: 7.3--7.8~eV~\cite{Mon02} and 6.6~eV~\cite{Cha82}.  Again,
we argue that this possible mismatch is no hindrance for the present and
qualitative exploration.

It is finally worth mentioning that, even if $\omega_1$ and $\omega_2$
are below IP, there is a basic difference of C$_3$ compared with
the Mg atom and the ${{\rm K}_9}^+$ cluster~: A still
  significant part of the dipole strength lies above IP. These excitations reside
  in the electron continuum and emit directly, thus overlaying the
  signal from the (discrete) resonances. This indicates that a PES
  from an instantaneous boost might be more involved than in the two previous
  examples. This will be confirmed in the next section.

\subsection{Multi-mode excitation peaks in PES}

We now turn to the PES and PAD of C$_3$.  To avoid interference 
with transverse modes, we apply the
instantaneous boost in the longitudinal (or $z$) direction only. The
obtained PES are shown in the left panel of Figure~\ref{fig:pad-c3}
for three different boost momenta $p_z$, $p_0=0.05/$a$_0$, $1.6p_0$,
and $2p_0$.
\begin{figure*}[htbp]
 \centering
 \includegraphics[width=\linewidth]{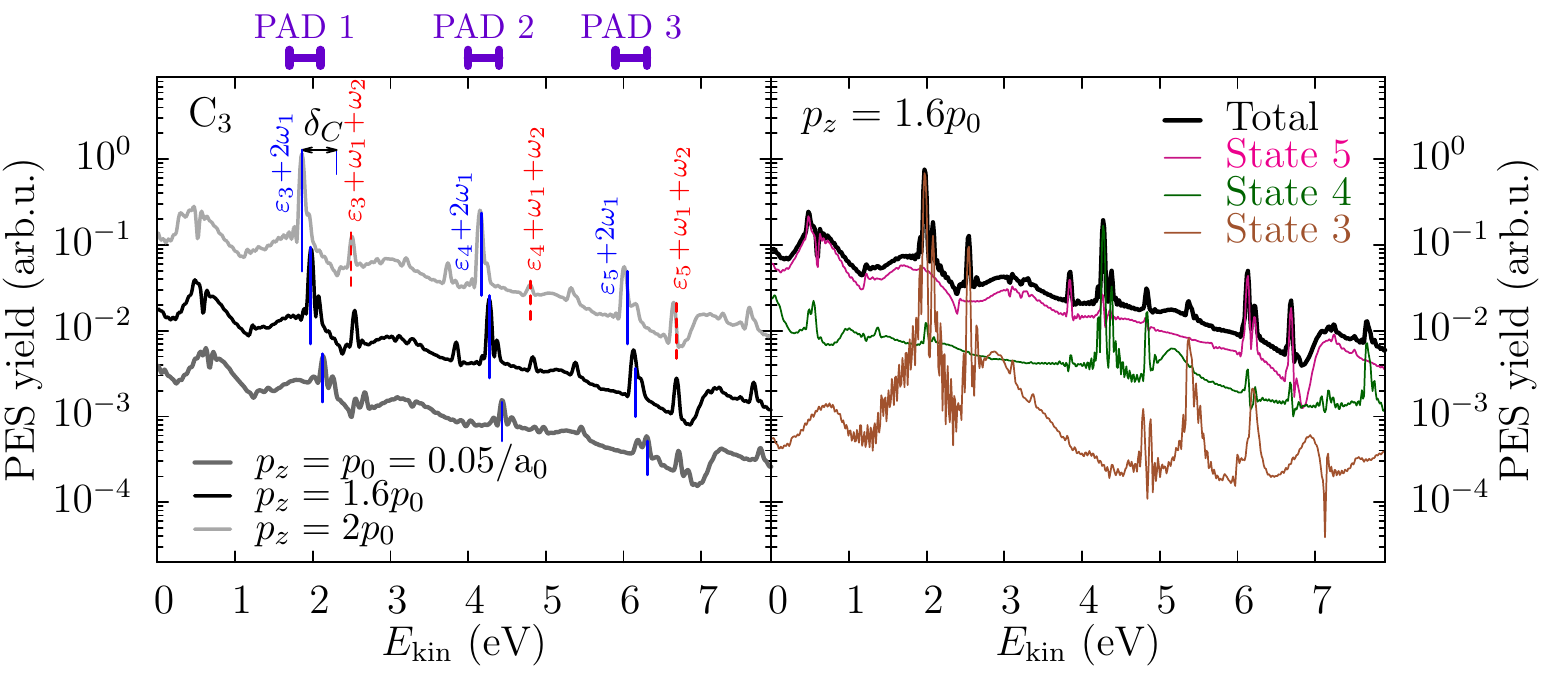}
 \caption{ Left~: Photoelectron spectra for C$_3$ excited by an
   initial instantaneous boost along longitudinal ($z$) direction with
   different boost strengths as indicated. Results are rescaled for
   better visualization. The vertical lines indicate expected $E_{\rm kin}$
   according to Eq.~(\ref{eq:multiplas-coul}) for
   $2\omega_1$-excitations (solid) and for
   $\omega_1+\omega_2$-excitations (dashes) of s.p. states labeled in
   Figure~\ref{fig:c3_stat-opt}. Here, $\delta_C$ indicates a Coulomb
   shift of the peaks (see text for details). The horizontal bars
   marked by PAD $i$ with $i=1,2,3$ at the top indicate the energy
   regions over which the corresponding PAD are plotted in Figure
   \ref{fig:pad-c3}.  Right~: State-resolved PES for boost
   $p=p_z=1.6p_0$.}
 \label{fig:pes-c3}
\end{figure*}
We clearly see strong peaks sticking out of a noisy background.  The
vertical lines indicate the peak positions predicted by the
rule~(\ref{eq:multiplas}), dashed lines for twice the $\omega_1$ mode
and dotted lines for a combination of $\omega_1$ and $\omega_2$.  The
assignments are corroborated by checking with the PES from specific
states 3, 4, and 5, shown in the right panel of
  Figure~\ref{fig:pes-c3}. One can clearly identify which state produces
  which peak. The peaks in the total PES (left panel) show a trend
  with boost strenth: they grow in height and are down-shifted
to lower $E_\mathrm{kin}$ with increasing boost. The down-shift,
  also called ``Coulomb shift'', had already been observed before in
connection with laser induced PES~\cite{Poh00}.  It is due to
  ionization. Note that electron emission increases with boost
  strength~: we find a total ionization $N_\mathrm{esc}=$0.018,
0.05, and 0.082 for $p_0=0.05$/a$_0$, $1.6p_0$, and $2p_0$ respectively.  This
ionization deepens the Coulomb potential as $-N_\mathrm{esc}e/r$ in
the course of the electron emission which, in turn, leads to the
Coulomb shift of s.p. energies.
Therefore, Eq.~(\ref{eq:multiplas}) must be modified to 
\begin{equation}
E_{\rm kin} = \varepsilon_j +  \sum_{k=1}^{M} \nu_k\omega_k - \delta_C \quad,
\label{eq:multiplas-coul}
\end{equation}
where $\delta_C=N_\mathrm{esc}e/R_\mathrm{syst}$ stands for the
Coulomb shift and $R_\mathrm{syst}$ is the relevant system radius.  As
is visible in Figure~\ref{fig:pes-c3}, a given peak is gradually
redshifted, that is, $\delta_C$ increases with $p_z$. We have
emphasized by an horizontal arrow the maximum value of
  $\delta_C$ (here, 0.44~eV) obtained for the strongest boost.

Besides the Coulomb shift, the peaks grow with increasing
boost. This indicates that they stem from non-linear effects,
actually quadratic in the excitation. The background remains rather
inert, except for the global upshift. This suggests that most of the
background stems from direct emission by continuum modes. The
many bumpy structures in the background reflect the fact that the
dipole spectrum above IP is still rather structured, see right panel in
Figure~\ref{fig:c3_stat-opt}. These bumps are rather broad which
confirms that they correspond to continuum states with large escape
width. In contrast, the peaks from $2\omega_1$ processes are very
narrow because truly bound modes are involved. Of course, for stronger
excitations, there might also be other non-linear combinations of
modes and some cross-talk to transverse modes. However, the spectral
density of such events is large and the strength is weak. These
secondary effects are probably dissolved in the background.

Finally, Figure~\ref{fig:pad-c3} displays the combined PES/PAD zoomed
onto the regions of the three prominent peaks as indicated in
Figure~\ref{fig:pes-c3}.
\begin{figure}[htbp]
 \centering
 \includegraphics[width=\linewidth]{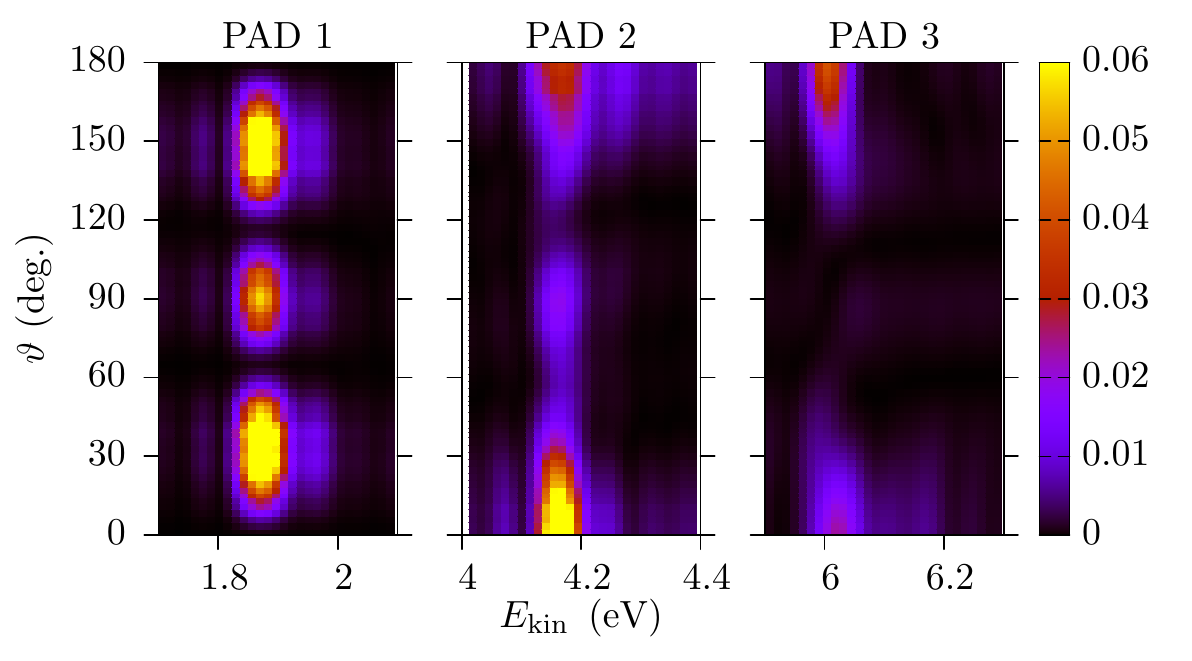}
 \caption{ Combined PES/PAD of C$_3$ for boost strength $p_z=0.05$/
   a$_0$ in the three energy ranges as indicated in Figure~\ref{fig:pes-c3}.
  }
 \label{fig:pad-c3}
\end{figure}
The 0 and 180$^\circ$ angles correspond to the longitudinal direction.
Although the excitation is done along this axis, electrons are not
emitted exclusively in this direction~: the deeper bound the state,
the less aligned the emission. Indeed, whereas PAD 3 (emission from
the HOMO) exhibits a maximum at $\theta=0$ and $\theta=180^\circ$, PAD
2 (emission from HOMO$-1$) possesses in addition a secondary maximum
in the transverse direction ($\theta=90^\circ$). And for PAD 1, the
emission is not aligned at all with the chain axis. It rather peaks at
$\theta=30^\circ$ and $150^\circ$, with a local maximum at
$90^\circ$. This appearance of state-specific sidewards emission
indicates that the processes considered here still stay in a regime of
moderate excitations. Further increased excitation will push emission
more to forward direction \cite{Poh04b}.  But this happens at the
price of washing out the PES, see Figure \ref{fig:K9p-plasmon}.

\section{Conclusions}

In this paper, we have explored the impact of strong
    resonance modes of an electronic system on the distribution of
    kinetic energies of emitted electrons, called Photo-Electron
    Spectrum (PES) in analogy to laser induced experiments.  Test
    cases are the Mg atom, ${{\rm K}_9}^+$ as a metal cluster, and
  the C$_3$ chain as a covalent molecule still exhibiting
  a partially metallic behavior.  To eliminate any frequency bias
  from outside, we employ a short-time excitation process through
  collision with a fast charged projectile. The Coulomb field
delivered by the bypassing ion is well described by an
instantaneous dipole boost of the electronic wave functions.
The resonant modes of the mo\-le\-cules have been determined by
  spectral analysis of the optical response in an independent
  calculation.

The resulting PES show sequences of pronounced peaks much
    similar to the case of laser excitation. But here the peaks have
  to be assigned as multi-plasmon (multi-resonance) excitations
  of the single particle states. Such an assignment is simple and
    obvious for the strongest resonance. It may be carried forth to
    secondary resonances if they are strong enough. Peaks with small
    strength and/or high spectral density disappear in the background
    signal. Henceforth, the conditions for the observation of these peaks
  are extremely demanding. One needs, first of all, a spectrum with
  one or two clearly dominating peaks which sets strong limits on
  possible systems. Even then, there is only a narrow window of
  excitation strength for the observation. For too weak an excitation
  (too large and impact parameter and/or too small a charge), the modes are not excited strongly
  enough to lift the peaks above background. For too strong an
  excitation (too small an impact parameter and/or too high a charge), the
  PES is smeared out as a consequence of the gradual downshift of the
  single particle spectrum. This also means that experimentally
the excitation of
  multiple-modes should be visible in a PES only for well selected sets of
  $b$,$Z$, and $v$.

\section*{Acknowledgements}
This work was supported by the 
Institut Universitaire de France, and was granted access to the HPC resources of IDRIS under the allocation 2014--095115
made by GENCI (Grand Equipement National de Calcul Intensif), of CalMiP (Calcul en Midi-Pyr\'en\'ees) 
under the allocation P1238, and of RRZE (Regionales Rechenzentrum Erlangen).


\end{document}